\documentclass[useAMS,usenatbib,usegraphicx]{mn2e}

\title[Magnetic and Metallic Degenerate]{The Magnetic and Metallic Degenerate G77-50}

\author[J. Farihi et al.]{J. Farihi$^1$,  P. Dufour$^2$, R. Napiwotzki$^3$,and D. Koester$^4$\\
$^1$Department of Physics \& Astronomy, University of Leicester, Leicester LE1 7RH, UK;
	jf123@star.le.ac.uk\\
$^2$D\'epartement de Physique, Universit\'e de Montr\'eal, Montr\'eal, QC H3C 3J7, Canada\\
$^3$Centre for Astrophysics Research, University of Hertfordshire, Hatfield AL10 9AB, UK\\
$^4$Institut f\"ur Theoretische Physik und Astrophysik, University of Kiel, 24098 Kiel, Germany}
\begin{document}

\date{}

\maketitle

\label{firstpage}

\begin{abstract}
An accumulation of multi-epoch, high-resolution, optical spectra reveal that the nearby star G77-50 
is a very cool DAZ white dwarf externally polluted by Mg, Fe, Al, Ca, and possibly Na, Cr, Mn.  The
metallic and hydrogen absorption features all exhibit multiple components consistent with Zeeman 
splitting in a $B\approx120$\,kG magnetic field.  Ultraviolet through infrared photometry combined
with trigonometric parallaxes yield $T_{\rm eff}=5310$\,K, $M=0.60$\,$M_{\odot}$, and a cooling 
age of 5.2\,Gyr.  The space velocity of the white dwarf suggests possible membership in the Galactic 
thick disk, consistent with an estimated total age of 8.6\,Gyr.  G77-50 is spectrally similar to G165-7 
and LHS\,2534; these three cool white dwarfs comprise a small group exhibiting both metals and 
magnetism.  

The photospheric metals indicate accretion of rocky debris similar to that contained in asteroids, 
but the cooling age implies a remnant planetary system should be stable.  A possibility for G77-50 
and similarly old, polluted white dwarfs is a recent stellar encounter that dynamically rejuvenated the
system from the outside-in.  Metal abundance measurements for these cooler white dwarfs have the 
potential to distinguish material originating in outer region planetesimals injected via fly-by.  If common 
envelope evolution can generate magnetic fields in white dwarfs, then G77-50 and its classmates may 
have cannibalized an inner giant planet during prior evolution, with their metals originating in terrestrial 
bodies formed further out.  Although speculative, this scenario can be ruled out if terrestrial planet 
formation is prohibited in systems where a giant planet has migrated to the inner region nominally 
engulfed during the post-main sequence.
\end{abstract}

\begin{keywords}
	circumstellar matter---
	planetary systems---
	stars: abundances---
	stars: evolution---
	stars: magnetic fields---
	white dwarfs
\end{keywords}

\section{INTRODUCTION}

Metal-enriched white dwarfs have experienced a resurgence of interest coinciding roughly with the 
launch of {\em Spitzer}.  Beginning with the landmark study of \citet{zuc03}, evidence has gradually 
accumulated supporting a picture whereby most, if not all cool, metal-lined white dwarfs obtain their 
photospheric heavy elements via rocky bodies similar in composition and in mass to large Solar 
System asteroids \citep{far10a,jur06}.  Data substantiating this picture comes primarily from {\em 
Spitzer} photometric detections of infrared excess indicating circumstellar dust orbiting within the 
Roche limit of a substantial fraction of metal-contaminated white dwarfs \citep{far09,von07}, and 
via mid-infrared spectroscopy of the disk material, revealing silicate minerals (primarily olivines) 
associated with rocky planet formation \citep{jur09a,rea05}.  

\begin{figure*}
\includegraphics[width=182mm]{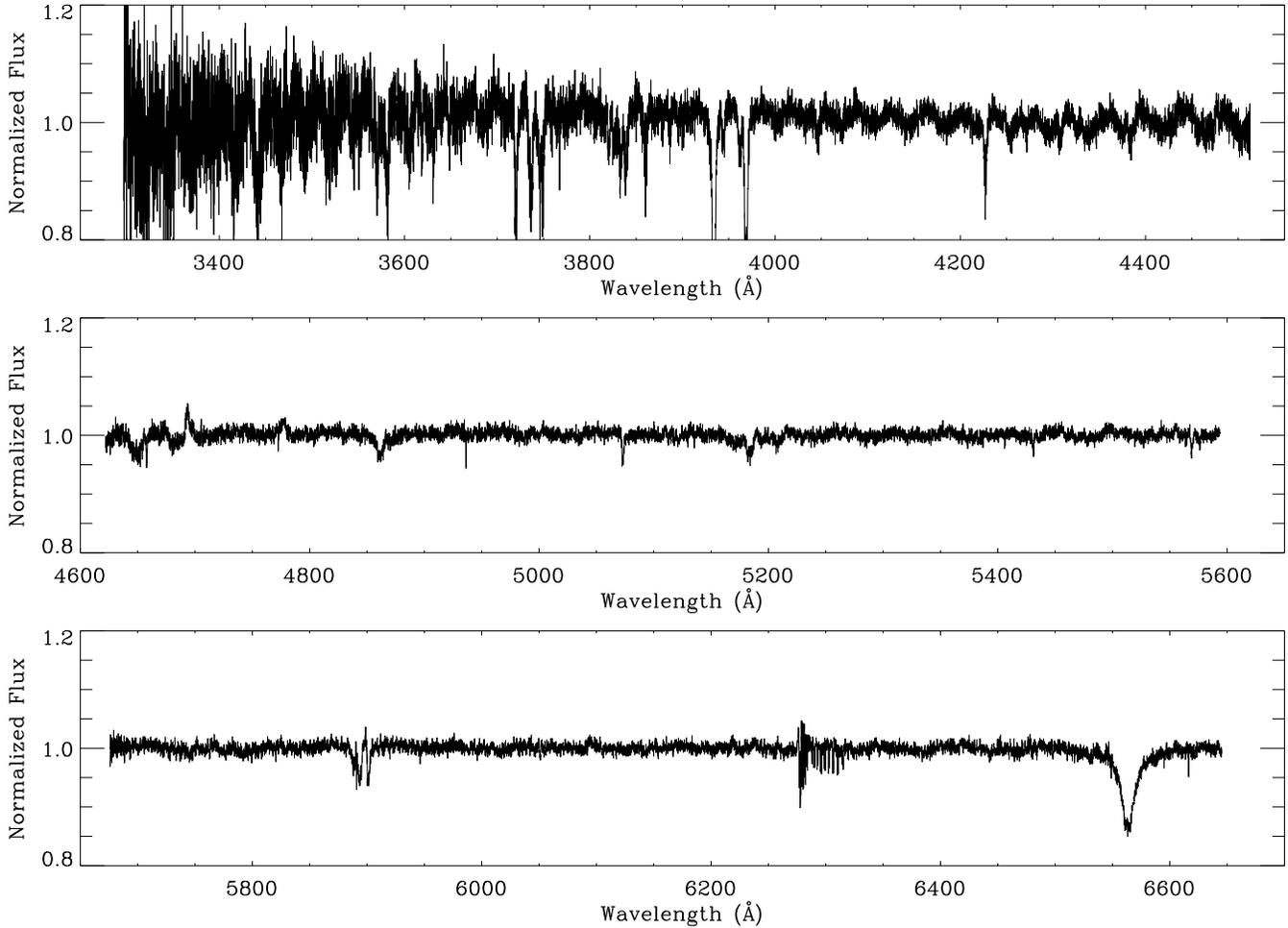}
\vskip -2 mm
\caption{The normalized and coadded UVES spectrum of G77-50, rebinned onto a grid with spacing 
0.1\AA.The quasi-periodic pattern prominent at shorter wavelengths is due to a light path difference 
between the flat-field lamp and the sky; this interference is not fully correctable, especially at high S/N.  
Several previously unidentified metal absorption features are evident, including lines of Mg, Fe, Al, Ca, 
and perhaps Na, Cr, Mn.  H$\beta$ is detected for the first time in a very cool white dwarf.  There are a 
large number of weak lines present in the spectrum whose authenticity is uncertain.  The features in the 
green arm below 4700\AA \ and those in the red arm near 5900\AA \ and 6250\AA \ are detector artifact 
and telluric absorption residuals (Table \ref{tbl1}).
\label{fig1}}
\end{figure*}

A powerful example of the scientific potential inherent in these stars is the spectacularly polluted 
white dwarf GD\,362 \citep{kaw05,gia04}.  This system shows remarkable infrared excess emission 
from closely orbiting dust, reprocessing over 3\% of the incident stellar flux and exhibiting perhaps 
the strongest silicate emission feature associated with {\em any} mature star \citep{jur07}.  The optical 
spectrum of GD\,362 displays 15 elements heavier than helium in an abundance pattern mimicking 
the Earth-Moon system \citep{zuc07}.  The convection zone of the star has been enriched, at a bare 
minimum, by the equivalent of a 240\,km asteroid, and possibly by a body as massive as Callisto or 
Mars, with promising evidence for internal water \citep{jur09b}.  Thus, white dwarfs enable unique 
insights into terrestrial planet formation around intermediate mass stars by providing data that can 
be obtained no other way; a lower mass limit to and the bulk chemical composition of extrasolar 
rocky, minor or possibly major planets.

This paper investigates the spectral, kinematical, and atmospheric properties of the white dwarf
WD\,0322$-$019 (G77-50).  The star has been known as a very cool and metal-lined degenerate for
over 35 years \citep{hin74}, but was only recently found to exhibit split Ca\,{\sc ii} H and K absorption
lines under high-resolution spectroscopy \citep{zuc03}.  Despite similar, high-resolution spectroscopic 
observations of several hundred white dwarfs \citep{koe09b,vos07}, including the detection of a few 
dozen with Ca\,{\sc ii} K absorption features \citep{koe05,zuc03}, only G77-50 and possibly LTT\,8381
display evidence for Zeeman-split metallic lines.  The observational program and scientific motivation 
are described in \S2, while the spectroscopic analysis, photospheric abundances, and stellar properties 
are derived in \S3.  The discussion of the results is contained in \S4, where potential origins for the 
metals and magnetism in G77-50 and related objects are explored in some detail.

\section{OBSERVATIONS \& DATA}

\subsection{Echelle Spectroscopy}

G77-50 was observed a total of 24 times between 2008 October 4 and November 25 at Cerro  
Paranal with the 8.2\,m Very Large Telescope Unit 2 using the Ultraviolet and Visual Echelle 
Spectrograph (UVES; \citealt{dek00}).  Spectroscopy was performed over the two detectors covering 
wavelengths from 3200\,\AA \ to 6650\,\AA\ using a standard dichroic configuration with $\lambda_c
=3900/5640$\,\AA, resulting in two narrow gaps in spectral coverage near 4550 and 5650\,\AA.  A slit 
width of $1\farcs0$ was employed with $2\times2$ binning, resulting in a nominal resolving power of 
$R\approx40\,000$ in both the UV-Blue and Red arms of the instrument.  The observations were taken 
at random intervals that were broadly logarithmic in temporal spacing, with each dataset consisting of 
two consecutive 900\,s exposures.  The featureless white dwarf WD\,0000$-$345 (LHS\,1008) was 
observed as a spectral standard on 2008 October 8 using an identical instrumental setup.

The echelle data were processed with the UVES pipeline version 4.3.0, including cosmic ray 
masking, flat fielding, wavelength calibration, order merging, and distilled using optimal aperture 
extraction.  Raw spectra produced in this manner had signal-to-noise (S/N) ratios that fell between 
16 and 22 at 5000\AA.  The spectral standard data were interpolated in wavelength to match the 
solution for the science target and smoothed by 121\,pixels (3.5 to 4.0\AA).  The science target 
was divided by the spectral standard and the resulting spectrum was multiplied by an appropriate 
temperature blackbody to achieve a relative flux calibration.  

The 2008 UVES dataset was supplemented by two archival UVES observations of G77-50 taken 
for the SPY survey in 2000 \citep{nap03}, and fully reduced Keck HIRES spectra obtained in 1999 
and 1998 \citep{zuc03}.  After correcting for heliocentric velocity at the time of individual exposures, 
a master spectrum was constructed from the normalized coaddition of all 26 UVES spectra.  The 
resulting spectrum has a bin width of 0.1\AA, which increases the S/N, while reducing the original 
resolution to approximately match the decrease resulting from the wavelength shifts observed in
the Ca\,{\sc ii} H and K lines (see \S3).  This spectrum is displayed in Figure \ref{fig1}, while Table 
\ref{tbl1} lists prominent stellar features and non-stellar artifacts.

\begin{table}
\begin{center}
\caption{Notable Features in the Coadded UVES Spectrum of G77-50\label{tbl1}} 
\begin{tabular}{@{}cccc@{}}
\hline

Wavelength	&Absorbing		&Wavelength		&Absorbing\\
(\AA)			&Element			&(\AA)			&Element\\

\hline

3441			&Fe\,{\sc i}		&3934			&Ca\,{\sc ii}\\
3570			&Fe\,{\sc i}		&3944			&Al\,{\sc i}\\
3582			&Fe\,{\sc i}		&3962			&Al\,{\sc i}\\
3619			&Fe\,{\sc i}		&3968			&Ca\,{\sc ii}\\
3631			&Fe\,{\sc i}		&4227			&Ca\,{\sc i}\\
3648			&Fe\,{\sc i}		&4600$-$4700		&...\\
3720			&Fe\,{\sc i}		&4775			&...\\
3735			&Fe\,{\sc i}		&4861			&H\,{\sc i}\\
3749			&Fe\,{\sc i}		&5073			&...\\
3758			&Fe\,{\sc i}		&5183			&Mg\,{\sc i}\\
3767			&Fe\,{\sc i}		&5890			&Na\,{\sc i}\\
3820			&Fe\,{\sc i}		&5900			&...\\
3832			&Mg\,{\sc i}		&6250$-$6350		&...\\
3838			&Mg\,{\sc i}		&6563			&H\,{\sc i}\\
3860			&Fe\,{\sc i}		&				&\\

\hline

\end{tabular}
\end{center}

{\em Note}.  Owing to magnetism, the line centers cannot be determined accurately and 
the wavelengths given are approximate or laboratory values.  In some regions, many lines of Fe\,{\sc i} 
contribute to a single, complex feature but only one wavelength is listed for simplicity.  Those features 
without a corresponding element are due (in part or in whole) to detector artifacts or telluric absorption 
residuals.

\end{table}

\section{SPECTROSCOPIC ANALYSIS \& SYSTEM PARAMETERS}

\begin{figure}
\includegraphics[width=86mm]{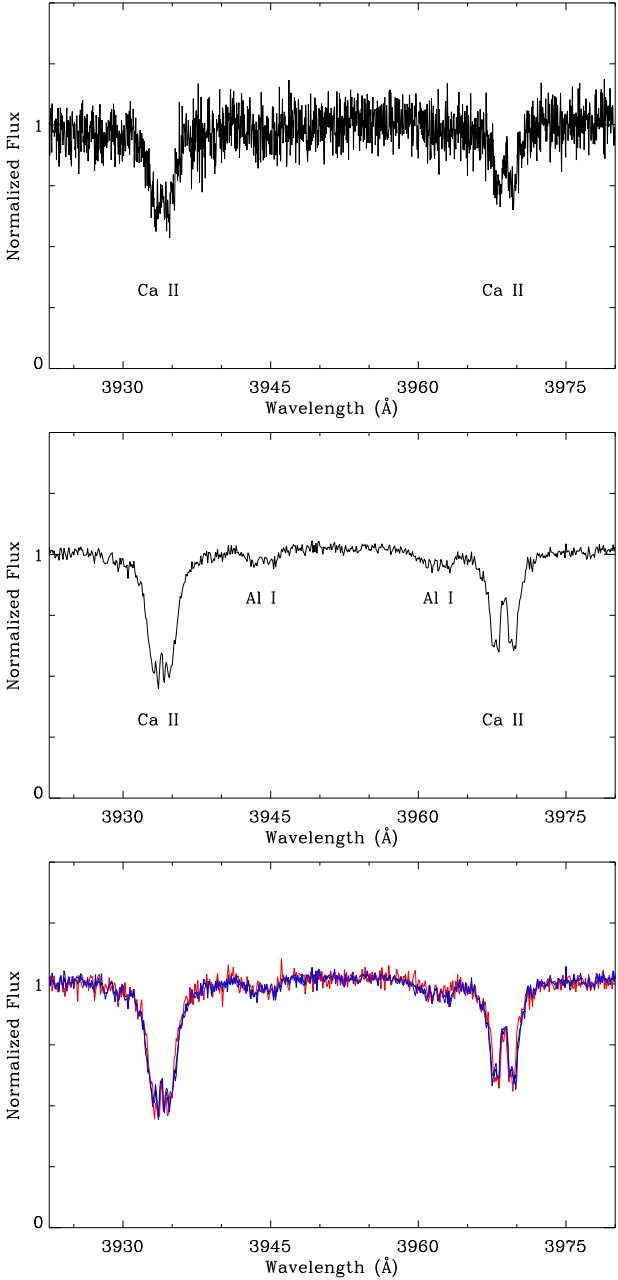}
\vskip -15 mm
\caption{{\em Upper}:  An example of an unbinned, single UVES spectrum of G77-50 in the 
Ca\,{\sc ii} H and K region.  Each individual exposure exhibits these double-peaked, metal absorption 
lines.  {\em Middle}:  The same region in the combined spectrum of all 26 UVES exposures.  In this deep 
spectrum, broad absorption features due to Al\,{\sc i} are revealed.  {\em Lower}:  In red and blue, the two
coadded spectral subsets discussed in \S3.1.1 are shown, plotted on top of the master spectrum in black.
\label{fig2}}
\end{figure}

G77-50 is listed as a binary suspect in \citet{ber97} based on the shape of its H$\alpha$ absorption 
feature and the lack of a trigonometric parallax distance at that time.  Seeming to confirm this hypothesis, 
\citet{zuc03} reported the white dwarf as binary based on what appeared to be double lines of Ca\,{\sc ii} 
H and K with large velocity separations, and possible changes in these velocities between two epochs of 
observation.  On the basis of this sound interpretation, the original aim of the UVES program reported here 
was to obtain the orbital period and mass ratio of the putative binary by measuring and monitoring the 
velocities of the Ca\,{\sc ii} H and K line components.  A typical, individual UVES spectrum of G77-50 is 
shown in the top panel of Figure \ref{fig2} and reveals the seemingly double Ca\,{\sc ii} features first 
detected by \citet{zuc03}.

\begin{table}
\begin{center}
\caption{Ca\,{\sc ii} H and K-line Velocity Measurements for G77-50\label{tbl2}} 
\begin{tabular}{@{}cccc@{}}
\hline

HJD$-$2\,400\,000	&$V_-$				&$V_+$				&Instrument\\
(d)				&(km\,s$^{-1}$)			&(km\,s$^{-1}$)			&\\

\hline
51158.87475		&$-43.79\pm1.45$		&$+82.10\pm1.83$		&HIRES\\
51404.10902		&$-51.36\pm3.11$		&$+99.77\pm3.61$		&HIRES\\
51737.93210		&$-61.34\pm5.43$		&$+94.93\pm6.89$		&UVES\\
51740.88365		&$-50.32\pm4.75$		&$+87.94\pm3.73$		&UVES\\
54743.71492		&$-43.26\pm2.69$		&$+98.32\pm2.32$		&UVES\\
54743.72596		&$-48.18\pm2.71$		&$+92.99\pm2.21$		&UVES\\
54743.73782		&$-46.78\pm2.78$		&$+96.59\pm3.60$		&UVES\\
54743.74887		&$-48.55\pm2.61$		&$+91.46\pm3.16$		&UVES\\
54743.81040		&$-50.28\pm2.87$		&$+91.03\pm2.64$		&UVES\\
54743.82149		&$-49.51\pm2.15$		&$+86.41\pm2.90$		&UVES\\
54744.73366		&$-50.01\pm1.79$		&$+89.36\pm1.82$		&UVES\\
54744.74471		&$-47.93\pm2.43$		&$+91.32\pm2.41$		&UVES\\
54744.83545		&$-51.48\pm1.99$		&$+89.90\pm2.40$		&UVES\\
54744.84656		&$-45.99\pm2.04$		&$+87.58\pm1.80$		&UVES\\
54747.69594		&$-54.15\pm2.32$		&$+97.44\pm2.32$		&UVES\\
54747.70700		&$-56.27\pm2.51$		&$+94.08\pm2.27$		&UVES\\
54749.65500		&$-54.65\pm4.22$		&$+88.54\pm2.95$		&UVES\\
54749.66605		&$-53.95\pm2.97$		&$+93.02\pm2.77$		&UVES\\
54750.66219		&$-58.50\pm1.87$		&$+91.74\pm1.96$		&UVES\\
54750.67321		&$-52.67\pm2.09$		&$+92.08\pm2.26$		&UVES\\
54759.63148		&$-46.90\pm3.15$		&$+81.53\pm3.08$		&UVES\\
54759.64259		&$-38.68\pm2.40$		&$+94.38\pm2.40$		&UVES\\
54760.73176		&$-44.87\pm2.32$		&$+79.17\pm2.55$		&UVES\\
54760.74281		&$-42.05\pm3.26$		&$+85.12\pm3.15$		&UVES\\
54767.68013		&$-44.27\pm2.07$		&$+76.15\pm2.15$		&UVES\\
54767.69118		&$-38.95\pm1.70$		&$+82.10\pm1.35$		&UVES\\
54795.63988		&$-41.59\pm2.30$		&$+81.30\pm2.55$		&UVES\\
54795.65095		&$-37.12\pm1.92$		&$+79.29\pm1.82$		&UVES\\
\hline

\end{tabular}
\end{center}
\end{table}

\subsection{Analysis and Interpretation of the `Double' Calcium Lines}

\subsubsection{Morphological Origins and Issues}

Initially, the two components seen in the Ca\,{\sc ii} H and K lines of each UVES spectrum were treated 
as originating in distinct stellar constituents in a spatially-unresolved binary.  New measurements taken 
in 2008 (MJD $>$ 54000) were combined with velocities obtained via identical analysis of archival UVES 
and Keck HIRES data to determine an orbital period.  Each major component of the Ca {\sc ii} features 
were simultaneously fitted by a Gaussian plus a Lorentzian line profile with the programme {\sf FITSB2} 
\citep{nap04}.  The parameters of the line profiles were determined simultaneously for all spectra, i.e.\
assuming a constant line strength and shape (more on this below).  The component velocities thus 
obtained are listed in Table \ref{tbl2}.  If the measured variations in the line shifts were due to binarity, 
then each stellar component should exhibit both red- and blue-shifted lines over time.  Under this 
assumption, a sinusoidal period of almost exactly 2.0\,d can be found within the line velocity data.

However, if the observed changes in the Ca\,{\sc ii} H and K line components are the result of orbital 
motion in two components of a binary, then the resulting velocity amplitudes predict an essentially 
equal mass system.  One can then estimate the radii and masses of the components based on two 
independent trigonometric parallaxes for G77-50; \citet{sma03} report $\pi=59.5\pm3.2$\,mas or 
$d=16.8^{+1.0}_{-0.9}$\,pc, while the US Naval Observatory (H. Harris 2010, private communication) 
measure $\pi=58.02\pm0.44$\,mas or $d=17.2^{+0.2}_{-0.1}$\,pc.  Taking the latter value by virtue of 
its smaller error, the predicted absolute magnitude of each equally luminous component would be
$M_V=15.7$\,mag.  At a nominal effective temperature of 5200\,K \citep{ber97}, this corresponds
to 0.9\,$M_{\odot}$ components.

\begin{figure}
\includegraphics[width=54mm,angle=-90]{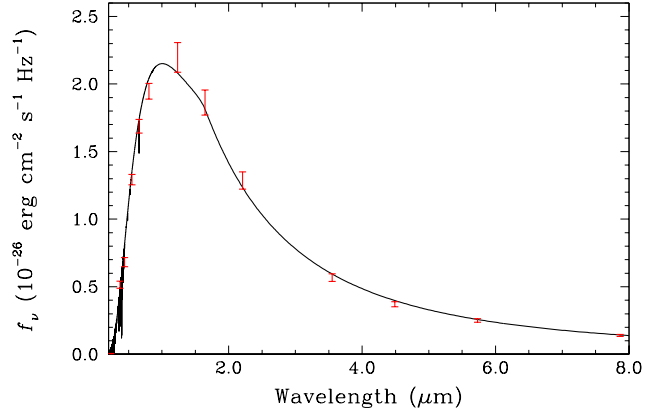}
\vskip 2 mm
\caption{Spectral energy distribution of G77-50 as revealed by {\em GALEX} \citep{mar05}, $UBVRI
JHK$ \citep{mcc99,ber97}, and {\em Spitzer} IRAC \citep{far08} photometry .  The red error bars plot 
the observational data, and the black solid line is the stellar atmosphere model with parameters listed
in Table \ref{tbl3}
\label{fig3}}
\end{figure}

While this possibility is exciting as the total mass would then easily exceed the Chandrasekhar 
limit, it also leads to a major inconsistency.  The ultraviolet through infrared photometric energy
distribution of G77-50 is consistent with a single effective temperature (Figure \ref{fig3}), and thus 
if the system were binary, the stars would be twins to all limits of observation, in temperature, in 
mass-radius, and also in metal abundance as the Ca\,{\sc ii} H and K line components are equally 
strong.  However, the profiles of the H$\alpha$ line in individual UVES spectra are found to be
inconsistent with hydrogen-rich white dwarf models with $\log\,g >$ 8.1; high surface gravities
fail to reproduce the observed line.  The results of the H$\alpha$ line fitting are consistent with 
a single, cool white dwarf at $d=17.2$\,pc as determined by trigonometric parallax.

Based on the discrepancies described above, the master spectrum was constructed as detailed in 
\S2 and carefully examined by eye.  The individual spectra exhibit only H$\alpha$, Ca\,{\sc ii} H and 
K prominently, but the combination of all UVES exposures yields a S/N $\approx100$ spectrum with 
many additional metal features and weak H$\beta$ absorption.  Importantly, all detected features in 
the coadded spectrum show evidence for multiple components, indicating the presence of a weak
magnetic field.  

Given the spread in line component velocities as shown in Table \ref{tbl2}, it is possible the lines will
be smeared or otherwise diluted by the coaddition of all individual UVES spectra.  To investigate the 
magnitude of this effect, the distribution of line velocities was searched for subgroups with similar values, 
whose coaddition might result in better line sensitivity.  Two groupings were identified and these spectral 
subsets were coadded as above: 7 spectra with $V_+$ in the range $75.5-84.5$\,km\,s$^{-1}$ and 15 
spectra with $V_+$ in the range $87.5-95.5$\,km\,s$^{-1}$.  A section of these two spectra are shown in 
the lower panel of Figure \ref{fig2} along with the master spectrum.  The noise level of the subset spectra
are increased compared to the master dataset, and there is no significant change in the shape or depth 
of the Ca\,{\sc ii} H and K lines, nor the weak Al\,{\sc i} lines.  [It is noteworthy that the moderately strong 
Ca\,{\sc i} line shown in the lower right panel of Figure \ref{fig4} compares favorably in sharpness to both 
the non-magnetic and Zeeman-split models (\S3.2).]  

\subsubsection{Constraints on Line Variability and Rotation}

In this alternative picture, the velocity variation observed in the Ca\,{\sc ii} H and K line components 
is most likely due to stellar rotation.  Assuming the star is not seen pole-on, the line of sight changes 
resulting from rotation span a (narrow) range of surface magnetic field strengths, and these induce 
changes in the strength of the observed Zeeman splitting.  

In order to search for periodicities in the wavelength shifts seen in the split H and K lines, a periodogram 
analysis was performed for 1) $\Delta V=|V_+-V_-|$ and 2) $V_+$ and $V_-$ independently.  The first
approach is adopted here, but the second yielded very similar results.  A best fit is found for $P=29.845
79$\,d and $\chi^2=25.9$, down from 280.5 when assuming constant line velocities, and thus indicating 
a high statistical significance of the variable solution.  The phase-folded velocity differences confirm a 
good quality fit with $\chi^2_{\rm red}=1.04$, with a number of somewhat poorer fits are present in the 
period range $28-33$\,d.  Two distinct solutions cannot be ruled out completely: 1.0311\,d ($\chi^2=33.8$) 
and 0.011333\,d ($\chi^2=35.6$).  The latter period is suspect as it is similar to the length of the individual 
UVES exposures, while periods near 1.0\,d are always dubious, and hence 29.85\,d is the most likely 
solution.  For a dipolar field configuration, half a rotation will produce a full cycle in the line velocities; 
if the long period solution is real, the white dwarf would be a slow rotator.

The relatively strong Ca\,{\sc ii} features were examined for variation in strength over time within the 
UVES dataset.  Measuring reliable equivalent widths for these blended lines was not possible, but the 
area swept out beneath the continuum was examined as a proxy.  Roughly half the feature strengths 
for both the H and K lines are found in a single, $0.5\sigma$-width peak, with all other measurements 
sitting below this peak, including some negative (i.e.\ unphysical) values.  Based on this analysis, the
individual spectra do not have sufficient S/N to meaningfully constrain variability in the observed line
strengths.

\subsection{Stellar Parameters and Abundances}

When first studied, H$\alpha$ was not detected in the optical spectrum of G77-50 and the white dwarf 
was typed as DG in the classification scheme of the period \citep{hin74}.  Subsequently, \citet{gre86} 
detected H$\alpha$ and \citet{sio90} typed the star in the modern spectral classification as a DZA white 
dwarf; strongest lines are metallic, weaker line(s) of hydrogen.  It is not clear whether feature strength 
should be determined by depth or equivalent width (E. M. Sion 2010, private communication), but in 
the case of G77-50, a DAZ classification is readily arguable as its atmosphere is hydrogen-rich, with 
trace metals.

As a first step, $T_{\rm eff}$ and $\log\,g$ are obtained by re-fitting existing $BVRIJHK$ photometric 
data with the latest pure hydrogen model atmosphere grid and using the new parallax measurement, 
following the method of \citet{ber97}.  From this, an effective temperature of $5310\pm100$\,K and a 
surface gravity of $\log\,g=8.05\pm0.01$ are obtained, corresponding to $0.60\pm0.01$\,$M_{\odot}$
and a cooling age of $5.2\pm0.1$\,Gyr.  Figure \ref{fig3} plots the global energy distribution obtained 
with these stellar parameters (listed in Table \ref{tbl3}), together with ultraviolet through mid-infrared 
photometry obtained from both the ground and space.  The consistency of the {\em Spitzer} data with 
the predicted stellar flux indicates excellent agreement with the model, which was fitted only at shorter 
wavelengths.

\begin{table}
\begin{center}
\caption{Stellar and Kinematical Parameters of G77-50\label{tbl3}} 
\begin{tabular}{@{}rccl@{}}
\hline

$T_{\rm eff}$					&&&$5310\pm100$\,K\\
$\log\,g\,({\rm cm\,s}^{-2})$		&&&$8.05\pm0.01$\\
$M$							&&&$0.60\pm0.01$\,$M_{\odot}$\\
$M_{\rm ms}$					&&&$1.4\pm0.2$\,$M_{\odot}$\\
$t_{\rm cool}$					&&&$5.2\pm0.1$\,Gyr\\
$t_{\rm ms}$					&&&$3.4\pm1.3$\,Gyr\\
$t_{\rm tot}$					&&&$8.6\pm1.3$\,Gyr\\
$\pi_{\rm trig}$					&&&$58.02\pm0.44$\,mas\\
$(\mu_{\alpha},\mu_{\delta})$		&&&$(241,-877)$\,mas\,yr$^{-1}$\\
$z_{\rm g}c$					&&&31.5\,km\,s$^{-1}$\\
$v_{\rm rad}$					&&&$-11.2$\,km\,s$^{-1}$\\
$v_{\rm tan}$					&&&$74.3$\,km\,s$^{-1}$\\
$v$ 							&&&75.1\,km\,s$^{-1}$\\
$(U,V,W)$ 					&&&$(-44.2,-60.5,-5.0)$\,km\,s$^{-1}$\\

\hline
\end{tabular}
\end{center}

\end{table}

\begin{figure*}
\includegraphics[width=132mm,angle=-90]{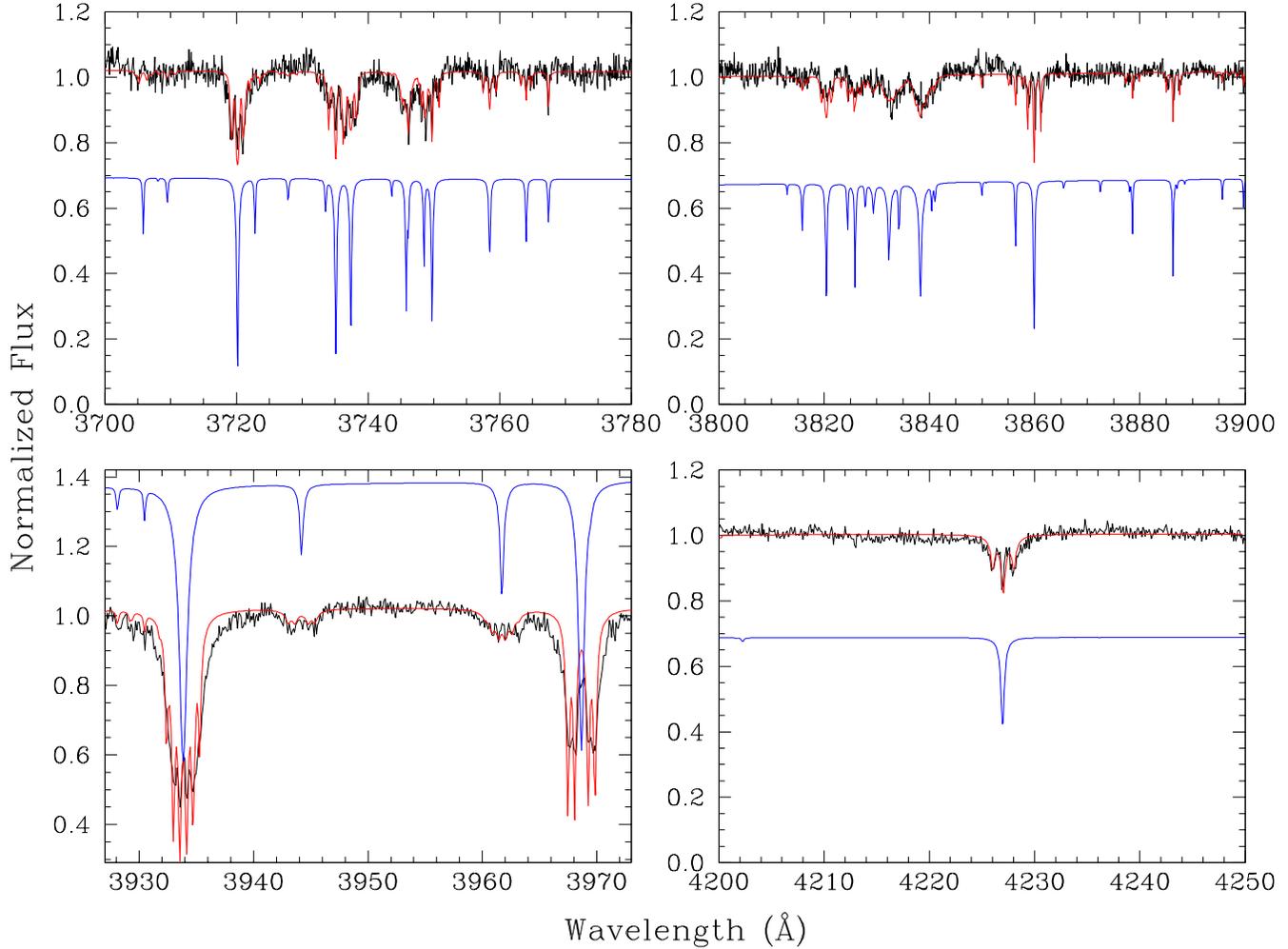}
\vskip 2 mm
\caption{The combined UVES spectrum of G77-50 is shown in black and the best-fit model in red.
Vertically offset from the data, the same best-fit, but non-magnetic model is shown in blue for line
identification.  The following wavelengths correspond to their non-magnetic positions.  {\em Upper
left}: All Fe\,{\sc i} lines.  {\em Upper right}:  Fe\,{\sc i} lines and Mg\,{\sc i} (3832,3838\,\AA).  {\em 
Lower left}: Ca\,{\sc ii} H \& K, Al\,{\sc i} (3944,3962\,\AA), and Fe\,{\sc i} (near 3930\,\AA).  {\em 
Lower right}: Ca\,{\sc i}. 
\label{fig4}}
\end{figure*}

\begin{figure*}
\includegraphics[width=132mm,angle=-90]{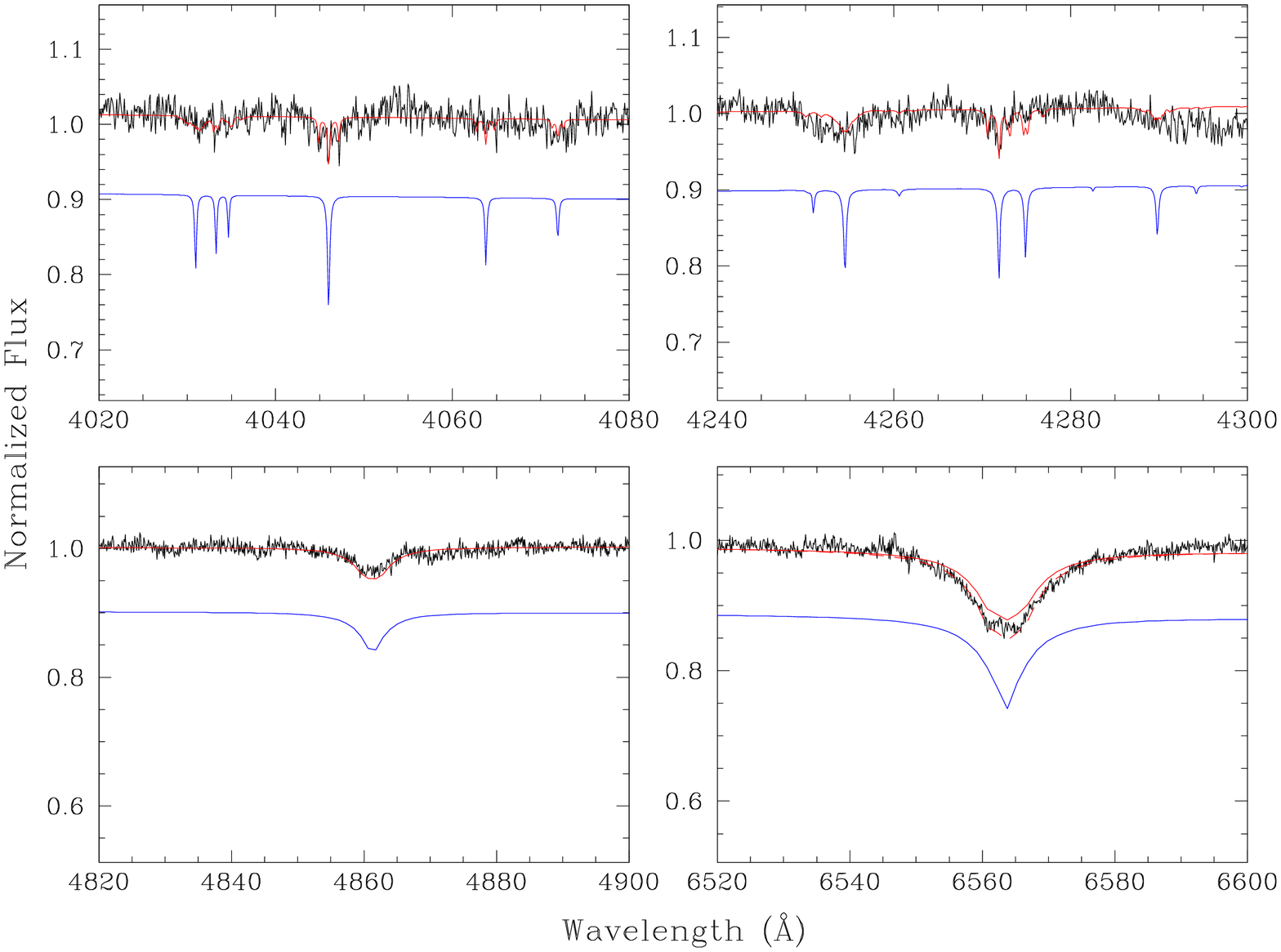}
\vskip 2 mm
\contcaption{{\em Upper left}:  Uncertain Mn\,{\sc i} features (4030,4033,4034\,\AA), and Fe\,{\sc i} 
lines.  {\em Upper right}:  Uncertain Cr\,{\sc i} features (4254,4274\,\AA), and Fe\,{\sc i} lines.  {\em 
Lower panels}:  Hydrogen Balmer lines $\alpha,\beta$.  The dashed line in the H$\alpha$ panel 
corresponds to a model 100\,K warmer than that used for the paper analysis (\S3.3).
\label{fig4b}}
\end{figure*}

The coadded spectrum of G77-50 indicates the presence of several elements as well as a magnetic 
field.  From the splitting of the various atomic lines, $B\sim120$\,kG is estimated.  Such a field is too 
weak to have an important effect on the thermodynamic structure of the star, and the absorption features 
are not strong enough to effect the determination of the atmospheric parameters described above.  This 
was verified explicitly {\em a posteriori} by comparing a pure hydrogen model energy distribution with 
that of a blanketed model that takes into account the presence of all the heavy elements at the derived 
abundances.  Hence, it is safe to use the above atmospheric parameters for the detailed abundance 
analysis presented below.  It is, however, necessary to include the effect of Zeeman splitting in synthetic 
spectra calculations to obtain accurate heavy element abundances.  As a first order approximation, 
magnetic line splitting is explicitly included in the synthetic spectrum calculation by assuming the 
linear Zeeman effect in a constant magnetic field of 120 kG over the surface of the star.

In the presence of a weak magnetic field, an atomic level with total angular momentum $J$ splits 
into $2J+1$ levels with magnetic quantum number $m = -J,...,+J$.  For each line, all transitions were 
calculated between the upper and lower levels that are allowed by the selection rule ($m_u - m_l= 0$,
+1, -1).  Each transition is then shifted by 

\begin{equation}
\Delta\lambda = 4.67\times10^{-13} \ \lambda_0^2 \ B \ (g_l m_l - g_u m_u) 
\end{equation}

\noindent
where $g$ is the Land\'e factor, $B$ the magnetic field strength in G, $m$ the magnetic quantum 
number of the level and $\lambda_0$ the central wavelength of the line in \AA.  In the few cases 
where the Land\'e factor is not given in the VALD\footnote{http://vald.astro.univie.ac.at} line list, the 
term designation is used to compute the Land\'e factor under the L-S coupling approximation, i.e.

\begin{equation}
g = 1 + \frac{J(J+1)+S(S+1)-L(L+1)}{2J(J+1)}
\end{equation}

\noindent
The relative strengths of the $\pi$, $\sigma$ ($\Delta m=0$, $\pm$1 respectively) components are 
calculated according to \citet{sob73}.
  
For the fixed value of $T_{\rm eff}$ and $\log\,g$ determined above, a grid of synthetic, magnetic 
spectra is then calculated for each of the strongly-detected metals, as well as for more uncertain 
elements.  The grids typically cover an abundance range from $\log\,[n({\rm Z})/n({\rm H})]=-8.0$ to 
$-11.0$ in steps of 0.5\,dex.  Next the abundance of each element is determined by fitting the various 
observed lines using a similar method to that described in \citet{duf05}.  The final adopted abundance 
of an element is taken as an average of all the fitted features, and these are listed in Table \ref{tbl4}.  
Finally, a synthetic spectrum including all the elements is calculated with the nominal abundances 
and plotted over the observed spectrum in Figure \ref{fig4}.  In order to facilitate the line identification 
of the various spectral features, a non-magnetic spectrum was also calculated with the same final 
metal abundances.

Despite the simplicity of using a constant 120\,kG field geometry for the synthetic spectrum calculations,
the fits are remarkably good.  Small discrepancies exist (e.g., in the wings and depth of the strong H \& K 
lines), that are probably an indication that the field strength varies slightly over the surface.  It should be
noted that the master, coadded spectrum represents an average taken at various, random rotation phases.  
Synthetic spectrum calculations using a more realistic (dipolar) magnetic field geometry, and a suitable 
average over the rotation period, would probably provide a better fit to the Ca\,{\sc ii} lines, but it is unlikely 
that such an effort would affect the derived abundances significantly.  

To find possible weak features arising from elements other than Mg, Fe, Al, and Ca, a synthetic spectrum 
was calculated with all elements up to Ni in abundances somewhat higher than their chondritic ratios 
\citep{lod03} relative to the observed Ca in G77-50.  This enhanced spectrum was laid over the coadded
spectrum and a careful search was done by eye, allowing identification of the strongest possible features
for additional elements as well as upper limits.  The search revealed the possible presence of very weak 
features of Mn\,{\sc i} (4030,4033,4034\,\AA) and Cr\,{\sc i} (4254,4274\,\AA; see Figure \ref{fig4}).  However, 
the putative features are only slightly above the noise level of the spectrum and thus these elements are 
considered uncertain.

The Na\,{\sc i} D doublet region also contains features (Figure \ref{fig1}), but caution is warranted as this 
portion of the spectrum is contaminated by detector artifacts which significantly complicate the analysis.  
A careful examination of all 26 individual UVES spectra indicates that the 5890\,\AA \ line is real and thus 
Na is likely detected in this star.  Unfortunately, the energy levels involved in these transitions are too 
close for the linear Zeeman approximation in a 120\,kG magnetic field, and the Paschen-Back regime
is appropriate.  Hence, the position and strength of the lines cannot be matched with the calculations 
presented here, and an accurate analysis of Na in this star is best left for future work with cleaner data.
Nevertheless, a rough Na abundance is obtained by matching the depth of the observed feature with 
the Zeeman approximation model, but interpretations based on this element should be avoided.

Lastly, the synthetic H$\alpha$ profile, appropriately split by a 120\,kG magnetic field, is found to be 
slightly weaker than the observed line, whose strength depends sensitively on the effective temperature
of the star.  A model with a temperature increased by 100\,K produces a line profile that matches the
observations well (Figure \ref{fig4}), and falls within the effective temperature uncertainty obtained 
from the fit to the photometric energy distribution.

\subsection{Stellar Kinematics}

Armed with an average, total velocity shift of $\gamma=+20.3$\,km\,s$^{-1}$ from the median of all 
the Ca\,{\sc ii} line centers in Table \ref{tbl2}, a mass-radius constrained by trigonometric parallax, the 
stellar radial velocity of G77-50 can be disentangled from its gravitational redshift.  The approximation

\begin{equation}
z_{\rm g} \approx \frac{G M}{c^2 R}
\label{eqn1}
\end{equation}

\noindent
is valid for white dwarfs.  For a star of 0.60\,$M_{\odot}$ and $\log\,g=8.05$, the radius is $R=0.0121$\,$
R_{\odot}$ and $v_{\rm g}=31.5$\,km\,s$^{-1}$.  Thus the radial velocity of the binary system is $v_{\rm 
rad}=\gamma-v_{\rm g}=-11.2$\,km\,s$^{-1}$.  Table \ref{tbl3} lists the resulting three dimensional space 
motion for G77-50 incorporating the above radial velocity and its observed proper motion \citep{zac10}.  
While the velocities are not extreme, the $V$ component lags behind the rotation of the Galaxy by 
60\,km\,s$^{-1}$, indicating the white dwarf belongs to the old thin or thick disk population \citep{pau06}.

A total age for the white dwarf can be estimated in the following way.  Taking the average and standard 
deviation of three independent initial-to-final mass relations \citep{wil09,kal08,dob06}, the main-sequence 
progenitor of G77-50 had a likely mass of $1.4\pm0.2$\,$M_{\odot}$.  Main-sequence lifetimes for stars
in this mass range are obtained from the analytical formulae of \citet{hur00}, and added to the cooling age
of the white dwarf.  This results in a total system age of $8.6\pm1.3$\,Gyr for G77-50, a range consistent
with thick disk membership, although not conclusively.

\begin{table}
\begin{center}
\caption{Metal Abundances and Masses in G77-50\label{tbl4}} 
\begin{tabular}{@{}lccc@{}}
\hline

Element 			&$\log\,[n({\rm Z})/n({\rm H})]$		&$M_{\rm z}$		&$t_{\rm diff}$\\
				&							&(10$^{20}$\,g)	&(10$^5$\,yr)\\

\hline

Na$^{\dag}$		&$-9.1$						&0.25			&3.76\\
Mg				&$-8.3$						&1.62			&3.85\\
Al				&$-9.2$						&0.20			&3.72\\
Ca				&$-9.8$						&0.08			&2.78\\
Cr$^{\ddag}$		&$-9.9$						&0.08			&2.49\\
Mn$^{\ddag}$		&$-10.3$						&0.03			&2.36\\
Fe				&$-8.7$						&1.31			&2.33\\
Total				&$-8.0$						&3.57			&\\

\hline

\end{tabular}
\end{center}

$^{\dag}$Abundance uncertain.

\smallskip
$^{\ddag}$Presence of this element is uncertain.

\smallskip
{\em Note}.  Errors are 0.2\,dex

\end{table}

\section{ORIGIN OF METALS AND MAGNETISM}

Some of the material presented in this section is necessarily speculative, and are first attempts to 
understand how a star such as G77-50 came to possess its somewhat unusual characteristics.  It 
should be the case that near-future theoretical and empirical investigations will be able to test these 
ideas more rigorously.  

\subsection{Nature of the Photospheric Metals}

Metal diffusion timescales and convection zone parameters for G77-50 were calculated using 
a complete model atmosphere with the parameters and heavy element abundances from the 
preceding analysis in the outer layers.  This was continued to the bottom of the convection zone using 
pure hydrogen \citet{koe09a}.  Table \ref{tbl4} lists the relevant lifetimes for all the metals detected or 
suspected in G77-50.  The total mass of the mixing layers in G77-50 is found to be $1.23\times10^{27}
$\,g, and can be combined with the photospheric abundances to obtain the current mass of each heavy 
element within the star.  As the metals continuously sink, these calculations yield the {\em minimum} 
mass of each accreted metal, and are listed in Table \ref{tbl4}.  The total mass of metals residing in the 
star is $3.57\times10^{20}$\,g and equivalent to that contained in a 65\,km diameter body of density 
2.5\,g\,cm$^{-3}$; the highest mass of heavy elements yet inferred within a hydrogen-rich white dwarf.

Owing to its very cool effective temperature, G77-50 possesses a sizable convection zone on par 
with those intrinsic to helium-rich white dwarfs \citep{koe09a}.  Thus, the star has a commensurately 
long diffusion timescale for heavy elements; e.g.\ 280,000\,yr for Ca.  In this and other respects, G77-50
is characteristic of the coolest metal-rich white dwarfs.  It has modest metal abundances, with $\log\,[n
({\rm Z})/n({\rm H})]<-8.5$ for all but Mg, a relatively low time-averaged accretion rate of $3.7\times10^
{7}$\,g\,s$^{-1}$ for all clearly detected and suspected heavy elements, and a lack of infrared excess 
as measured with {\em Spitzer} IRAC photometry \citep{far09,far08}, indicating a dearth or absence of
dust in its immediate circumstellar environment.

The relatively old white dwarf has a total space velocity of 75\,km\,s$^{-1}$ and its current location 
is 17.2\,pc from the Sun.  At this speed, the star travels nearly 77\,pc\,Myr$^{-1}$ and may be passing 
through the relatively interstellar matter-deficient Local Bubble \citep{red08,wel99}, its metal content
decaying from a past event within a dense region of gas and dust.  Within a period just over 1\,Myr,
G77-50 could have been outside the Local Bubble, and its metal abundances would have decayed
by a factor between 20 and 100, depending on the element.  With roughly $2\times10^{22}$\,g of 
atmospheric metals in this picture, how dense an interstellar region is necessary to account for the
accretion in such a manner?  For interstellar heavy elements typically contained in dust particles, the 
accretion rate onto a star is \citep{far10a}:

\begin{equation}
\dot M_{\rm z} = \frac{\pi G M  R \rho_{\infty}}{s} \left( \frac{T_{\rm eff}}{T_{\rm ev}} \right)^2
\label{eqn2}
\end{equation}

\noindent
where $M$ and $R$ are the stellar mass and radius, $s=\sqrt{v^2 + {c_s}^2}$ is the space velocity 
$v$ in the supersonic regime ($v\gg c_s$) appropriate for G77-50, and $T_{\rm ev}$ is the evaporation
temperature of the dust grains.  Conservatively estimating that dust evaporates at 1000\,K, the minimum
density required to deposit $2\times10^{22}$\,g of metal in the star over a typical sinking timescale of $3
\times10^5$\,yr is 1600\,cm$^{-3}$.  While such densities exist in molecular clouds, the space velocity of
G77-50 requires continuous accretion over a distance of 30\,pc in order that sufficient material is captured.
This scenario is unlikely; one of the nearest and largest cloud complexes, the Orion Nebula is only about 
8\,pc in diameter.

Alternatively, the metals in the star were accreted from its circumstellar environment and were 
originally contained within remnant planetary bodies rich in heavy elements.  The detection of both
Al and Ca in the star supports a scenario in which the metals originated in a refractory-rich source
such as a rocky, minor or major planet.  This possibility is likely based on the weight of evidence 
favoring circumstellar accretion among the observed population of single, cool, and metal-enriched 
white dwarfs \citep{zuc10,far10a}.  In this scenario, the photospheric metals in G77-50 were delivered 
by one or more rocky planetary bodies passing close enough to the white dwarf to become tidally 
destroyed or otherwise accreted \citep{jur03}.  

\begin{table}
\begin{center}
\caption{Metal Ratios in G77-50 and Other Polluted White Dwarfs with Al\label{tbl5}} 
\begin{tabular}{@{}llcccr@{}}
\hline

Star				&Name		&Mg/Fe			&Al/Fe			&Ca/Fe			&Disk?\\
\hline

0322$-$019		&G77-50		&2.9				&0.31			&0.08			&$-$\\
				&			& \ 1.2$^{\dag}$	& \ 0.14$^{\dag}$	& \ 0.06$^{\dag}$	&\\

\hline

				&solar		&1.2				&0.10			&0.07			&$-$\\
0208$+$096		&G74-7		&1.0				&0.08			&0.11			&$-$\\
0300$-$013		&GD\,40		&1.7				&0.21			&0.40			&$+$\\
1633$+$433		&G180-63		&2.1				&0.14			&0.18			&$-$\\
1729$+$371		&GD\,362		&0.5				&0.18			&0.26			&$+$\\

\hline

\end{tabular}
\end{center}

$^{\dag}$At the end of an {\em assumed} accretion event 0.5\,Myr prior.

\smallskip
{\em Note}.  The three middle columns list the relative number abundances.  Metal ratios for 
other white dwarfs with Al detections are taken from the literature \citep{kle10,zuc07,zuc03,lod03}.

\end{table}

For a prior pollution event, the current metal-to-metal ratios have been altered from the accreted 
values, via the individual heavy element sinking timescales.  Table \ref{tbl5} lists the photospheric 
abundance ratios, relative to Fe, for the four heavy elements confidently identified in G77-50.  Notably, 
both Mg/Fe and Al/Fe appear enhanced relative to solar values and to the stars with circumstellar dust; 
precisely as expected in a declining phase because Fe sinks most rapidly.  Interestingly, Ca/Fe is nearly 
solar, yet down by a factor of a few compared with the stars currently accreting from disks.  Under the 
questionable assumption that the accreted abundances were close to solar, then the photospheric 
pollution halted roughly 0.5\,Myr ago.  In all likelihood, the accretion epoch for G77-50 ended within 
the last few to several diffusion timescales, no longer than a few Myr ago.  Events older than 3\,Myr 
ago imply an Fe-dominated ($M_{\rm Fe}/M_{\rm z}\geq94$\%) parent body more massive than Pluto.  

\subsection{A Late Instability Near 5\,Gyr}

Again, G77-50 epitomizes the older and cooler, $T_{\rm eff}<9000$\,K metal-rich white dwarfs: 
something weighty has occurred recently in these Gyr old, presumably stable systems.  Generally,
the timescale for a planetary system to dynamically settle -- pre- or post-main sequence -- should be 
100\,Myr \citep{deb02}, and thus a catastrophic event at Gyr epochs is not expected for G77-50 and 
similar white dwarfs.  Dual planet interactions will occur during this short period, if they occur at all,
but instabilities among three (or more) planets may occur on longer timescales, as hypothesized for
the period of Late Heavy Bombardment in the Solar System \citep{gom05}.

Following \citet{deb02} and \citet{cha96} for the simple case of three 0.001\,$M_{\odot}$ planets, one 
can calculate system architectures that give rise to an instability timescale of 5.2\,Gyr during the white 
dwarf evolutionary stage.  Taking initial and final stellar masses to be 1.8 and 0.6\,$M_{\odot}$ produces 
an (adiabatic) orbital expansion factor of 3 between the main and post-main sequence phases.  For an 
innermost planet now located at 10\,AU, the outer planets would be near 30 and 157\,AU, with original 
-- and presumably metastable -- semimajor axes of 3.3, 10, and 52\,AU \citep{deb02}.  In general, for all 
layouts that keep the innermost planet safely outside the stellar photosphere ($a_1>2$\,AU) during the 
asymptotic giant branch, the planets must be spaced such that $a_2/a_1\ga3$ and $a_3/a_2\ga10$.
Thus, the onset of instability at Gyr timescales requires a somewhat finely-tuned, increasingly wide
planet spacing. 

An alternate way in which an old and stable planetary system might be efficiently agitated is a stellar 
encounter.  Though rare, at the cooling age and space velocity of G77-50, it can be shown that at least 
one fly-by is consistent with theoretical expectations.  Dynamically, the number of stellar encounters per 
unit time, within a distance $D$ of a star is given by \citep{gar99}:

\begin{equation}
N = \pi D^2 v_* \rho_*
\label{eqn3}
\end{equation}

\noindent
where $v_*$ is the space velocity of the white dwarf relative to passing stars, and $\rho_*$ is 
the local density of stars and stellar systems.  From {\em Hipparcos} data, the velocity dispersion of 
stars within 2\,kpc of the Sun and $|b|>30\,\degr$ varies between 22 and 44\,km\,s$^{-1}$ depending 
on spectral type \citep{mig00}.  Later type stars tend to be kinematically more stirred as they represent,
on the whole, relatively older populations.  And because the most abundant stars in the Galaxy are K 
and M dwarfs, these relatively fast moving stars are the objects for which encounters are most likely.
Taking 40\,km\,s$^{-1}$ for a typical field star and the 75\,km\,s$^{-1}$ space velocity of G77-50 yields
$v_*=85$\,km\,s$^{-1}$.  The local space density of stars is 0.081\,pc$^{-3}$ (T. Henry 2010, private
communication; \citealt{hen06})\footnote{Statistics of the solar neighborhood are continually updated 
at http://www.recons.org}, and Equation \ref{eqn3} gives $N=22$\,Myr$^{-1}$ for encounters within 1\,pc.
For close fly-bys that might significantly impact the outer regions of a planetary system, take $D\leq1000
$\,AU and then $N=0.5$\,Gyr$^{-1}$.  If correct, and such an encounter is 50\% probable within each 
0.5\,Gyr window, then it is 99.9\% likely that G77-50 has suffered at least one such stellar encounter 
within its 5.2\,Gyr cooling history.

An encounter could disturb a Kuiper Belt analog originally orbiting at a few to several tens of AU, but 
now residing at one to a few hundred AU owing to mass lost during the post-main sequence.  Large 
objects such as Sedna, whose mass is similar to Ceres, can readily be perturbed into high eccentricity 
orbits by a stellar encounter, thus sending outer planetesimals towards the inner system \citep{ken04}.  
While chances are tiny that any single, distant body could be flung directly within the tidal disruption 
radius of the white dwarf, the perturbation of many bodies (i.e.\ a significant fraction of a surviving 
population) could result in their capture and further scattering within the inner system \citep{deb02}.

A potential hurdle for this hypothesis is the total available mass that survives heating during the 
giant phases at these orbital distances.  Out to roughly 100\,AU, objects composed of pure water ice 
and up to 100\,km in diameter should sublimate completely \citep{jur04,ste90}, but the overall effect 
on realistic bodies of heterogenous chemical composition is unknown.  For example, some models 
predict that subsurface volatiles should be protected by superior layers of non-volatile material such 
as silicates \citep{jur10}.  Studies of Kuiper Belt analogs in the post-main sequence may reveal the
overall impact of sublimation and constrain the total mass that survives to the white dwarf stage 
\citep{bon10}.

In contrast, Oort cloud analogs will not be destroyed via heating but are prone to dynamical 
evaporation in the post-main sequence if the stellar mass loss is asymmetric \citep{par98}.  At tens 
of thousands of AU, stellar encounters can be frequent over Gyr timescales typical of very cool white 
dwarfs such as G77-50.  These events should strip away some portion of any Oort-like cloud, but also
perturb another fraction onto eccentric orbits overlapping with the inner system.  If such cold, outer
planetesimal belts are ultimately responsible for some fraction of metal-polluted white dwarfs, then a 
large deposition of volatile elements would be expected, but has not yet been observed \citep{zuc07,
jur06}.

The subset of older and cooler metal-rich white dwarfs that G77-50 represents are a relatively high
velocity group of stars, typically with $T>50$\,km\,s$^{-1}$ \citep{aan93}, and the above exercise can
be broadly applied to Gyr age white dwarfs with metals.  Thus, stellar encounters have the potential 
to make ancient planetary systems dynamically young for a brief period, and may account for the 
population of very cool DAZ and DZ stars, of which vMa\,2 is the prototype.

\subsection{Weak Magnetism}  

In the favored model for cool, metal-rich white dwarfs, a planetary system has survived post-main 
sequence evolution \citep{deb02}.  Instabilities drive rocky planetary bodies such as asteroids into 
close approach with the compact star, where they become tidally- or otherwise destroyed and are 
sometimes observed as dust disks \citep{jur03}.  For the high eccentricity orbits necessary for the
(eventual) delivery of metals onto the surface of the star, conventional sized planets are the most 
efficient perturbers.  In this picture, an otherwise-replete planetary system persists at metal-enriched 
white dwarfs, likely truncated near the maximum radius of the asymptotic giant progenitor star.

Based on precision radial velocity planet searches, 10.5\% of FGK dwarfs with masses in the range 
$0.7-1.3$\,$M_{\odot}$ host gas giant planets within 3\,AU \citep{cum08}.  Due to the finite time period 
over which these searches have been carried out, most of these giant planets orbit within 1\,AU, and 
a decent fraction orbit close to their host star (the so-called hot jupiters).  Relative to these findings for 
solar-type stars, giant planets are found more frequently, and with a higher mass distribution, in radial 
velocity searches of intermediate mass stars.  Both \citet{lov07} and \citet{jon07} find that planet frequency 
and planet (minimum) mass scale with stellar host mass, and that the fraction of gas giants orbiting within
3\,AU of stars with masses between around 1.5 and 2.0\,$M_{\odot}$ is closer to 25\% \citep{bow10}.  
Interestingly, stars with higher masses do not host closely-orbiting giant planets, possibly due to halted 
orbital migration from rapid inner disk dissipation \citep{cur09}, or planet-star gravitational tides 
\citep{han10}.

The bulk of giant planets orbiting within a few AU of their host stars will be destroyed during the 
post-main sequence phases of their host stars.  During the first ascent giant branch (RGB), stars 
with main-sequence masses above 1.0\,$M_{\odot}$ expand to around $10-20$\,$R_{\odot}$, or 
$0.05-0.10$\,AU, sufficient to swallow only the hot planets \citep{bow10,vil09}.  Depending on which
initial-to-final mass relation one uses, the main-sequence progenitor of G77-50 was an early to late 
F star with a mass between 1.2 and 1.6\,$M_{\odot}$.  For a $M>1.3$\,$M_{\odot}$ progenitor, there
is a 25\% chance it hosted a giant planet within a few AU, but not a hot planet.  In this case, the inner
planet would survive the RGB intact but most likely become enveloped during the asymptotic giant 
branch (AGB), either directly or via tidal forces; a 1.5\,$M_{\odot}$ star has a maximum AGB radius 
near 2.4\,AU \citep{vil07}.

Depending on the mass and angular momentum of an engulfed planet, a common envelope may
develop for a time sufficient to generate a magnetic dynamo \citep{nor10,tou08,sie99b,reg95}.  The 
eventual accretion of the planet onto the stellar core may also result in enhanced mass loss from
the giant \citep{sie99a}, producing a white dwarf remnant potentially less massive than predicted by 
initial-to-final mass relations.  If this occurs at the level of 0.05\,$M_{\odot}$ in the white dwarf, then 
the inferred mass of the progenitor star will be biased towards lower, main-sequence masses.  In 
the model of \citet{pot10}, the longer the common envelope environment, the larger the magnetic 
field strength in the resulting white dwarf plus companion merger.  This model predicts fields up to 
10$^7$\,G in a common envelope lasting roughly $10^4$\,yr, but with a stellar companion generating 
the convective motions.  Following \citet{pot10} but reducing the companion mass by a factor of 100 
(to 10\,$M_{\rm J}$), all else being equal, a magnetic field of 10$^5$\,G is predicted.  Interestingly, 
the surface field of G77-50 is $1.2\times10^5$\,G.  

\subsection{Possible Planetary System Architecture}

In the admittedly speculative picture outlined above, the primordial G77-50 system formed a gas 
giant that migrated through a disk to an inner orbit.  Subsequently and beyond this orbit, additional 
planets may have formed \citep{man07}.  As in the Solar System and a significant fraction of main-$
$sequence stars \citep{car09,su06}, a Kuiper belt analog existed at G77-50, consisting of primordial 
icy and rocky bodies condensed from the stellar nebula.  During post-main sequence evolution, the 
volatile outer layers of these planetesimals were lost by evaporation and radiation pressure, leaving 
behind objects with non-volatile surfaces and intact cores \citep{bon10,jur10}.

In order that a stellar encounter repopulate the inner, circumstellar regions around any star, sizable 
planets must exist there to trap or further scatter injected bodies.  Such captured planetesimals are 
vulnerable to perturbations via newly established resonances on 100\,Myr timescales \citep{deb02}.  
From this point forward in time, the eventual pollution of the star by a tidally destroyed planetesimal 
proceeds as envisioned for metal-rich white dwarfs with cooling ages less than 0.5\,Gyr \citep{far09,
jur03}.  If the stellar encounter hypothesis applies to G77-50 and other very cool, polluted white 
dwarfs, then the atmospheric metals may be a reflection of the rocky remains of primitive, outer system 
planetesimals rather than Solar System asteroid analogs.  As such, they would have a composition 
distinct from objects that formed at higher temperatures in the inner system, presumably including a 
higher concentration of volatile elements.

Testing this hypothesis may be difficult.  While cooler white dwarfs have lower atmospheric opacities 
and hence detection of their heavy elements is less challenging, there are two complications.  The main 
difficulty is that the coolest DAZ and DZ stars are not amenable to ultraviolet spectroscopy, which is the 
ideal region to search for volatiles (or any element) in low abundance.  For example, trace amounts of 
atmospheric carbon are only seen in the optical spectra of cool, helium-rich white dwarfs, and only at 
abundances $\log\,[n({\rm C})/n({\rm He})]>-7$ \citep{duf05}.  If C/Fe $=8.3$ in G77-50 as in chondrites,
then it would have $\log\,[n({\rm C})/n({\rm H})]=-7.8$ and be undetectable in this manner.  Also, it is not
clear what fraction of volatile material would remain after heating and ablation during the post-main 
sequence \citep{jur06}.

An additional complication is that it is difficult to constrain the epoch of accretion in these cooler, 
metal-polluted stars.  Observationally, it is rare that cool DAZ or DZ have dust disks, and this may be the 
result of a significant reduction in the ratio of disk lifetime to cooling age.  The observed abundances can 
be used to constrain the epoch of accretion, but this is fundamentally biased towards the composition of 
Solar System objects and may not accurately reflect those of extrasolar planetary bodies.  This latter 
concern may be partially alleviated in the growing number of DZ stars being found in the Sloan Digital 
Sky Survey \citep{duf07}; follow up of these stars will better constrain the relative frequency of disks at 
older, metal-enriched stars and their abundances will hopefully provide clues to the nature of the 
polluting bodies.

\section{CONCLUSIONS}

The cool white dwarf G77-50 is found to be simultaneously metal-contaminated and magnetic.  
A hypothesis that may account for both properties is an evolved planetary system.  While highly 
uncertain, it is possible that G165-7 \citep{duf06} and LHS 2534 \citep{rei01} acquired their magnetic 
fields and photospheric metals in a manner analogous to G77-50.  In this tentative picture, an inner 
giant planet was swallowed by the post-main sequence progenitor, creating a common envelope 
whose convective motions generated the extant magnetic field.  Surviving rocky planetesimals have 
polluted the star with several refractory metals that are typical major constituents of asteroids and inner 
Solar System planets \citep{all95}.  However, with a cooling age of 5.2\,Gyr, it is difficult to imagine 
a planetary system architecture with instabilities sustained over such a period.  A plausible model 
is a stellar encounter that perturbs outer planetesimals into the inner system, where planets scatter
and trap them with a renewed dynamical potential for close approaches with the white dwarf.

The exact nature of the parent body whose metals currently pollute G77-50 will likely remain 
uncertain for the foreseeable future.  The white dwarf has been observed over a large wavelength 
range with the two most powerful, high-resolution optical spectrographs currently available, and 
with the exception of Na, the detections and abundances here are not likely to be improved upon.
Furthermore, the effective temperature of the star means that there is little flux to observe with the 
ultraviolet spectrographs on {\em HST}.  G77-50 and other very cool DAZ and DZ stars will have 
to await more powerful long wavelength observations with {\em JWST} or ALMA.

\section*{ACKNOWLEDGMENTS}

The authors thank the anonymous referee for helpful comments that improved the manuscript.
J. Farihi thanks I. N. Reid and B. Zuckerman for sharing their HIRES datasets, and H. Harris (and 
the USNO) for providing the details of their latest parallax results for G77-50.  P. Dufour is a CRAQ 
postdoctoral fellow.  This work is based on observations made with ESO Telescopes at Paranal 
Observatory under programme 382.D-0804(A).

\label{lastpage}

\end{document}